\documentclass[8pt,twocolumn]{article}
\usepackage{graphicx}
\usepackage[english]{babel}  
\usepackage[utf8]{inputenc} 
\usepackage{amsmath,amssymb}
\usepackage{bbm}
\usepackage{authblk}
\usepackage[a4paper, total={7in, 10in}]{geometry}

\newcommand{\ie}{\textit{i.\,e.}}

\newcommand{\ic}{i}

  \newcommand{\B}{\boldsymbol}
\newcommand{\tens}[1]{\B{#1}}

\newcommand{\rot}{\B{\mathrm{\nabla\times}}}

\newcommand{\Mop}{\mathcal{M}}

\newcommand{\ddroit}{\mathrm{d}}

\newcommand{\bra}{\ensuremath{\left\langle}}
\newcommand{\ket}{\ensuremath{\right\rangle}}
\newcommand{\mb}{\ensuremath{\,\middle| \,}}

\newcommand{\epstens}{\tens{\varepsilon}}

\newcommand{\epsA}{\epstens_{\rm a}}
\newcommand{\epsB}{\epstens_{\rm b}}
\newcommand{\epsC}{\epstens_{\rm c}}
\newcommand{\epsU}{\epstens_{\rm u}}

\newcommand{\mutens}{\tens{\mu}}

\newcommand{\muA}{\mutens_{\rm a}}
\newcommand{\muB}{\mutens_{\rm b}}
\newcommand{\muC}{\mutens_{\rm c}}
\newcommand{\muU}{\mutens_{\rm u}}

\newcommand{\depsA}{\Delta\epstens_{\rm A}}
\newcommand{\depsB}{\Delta\epstens_{\rm B}}
\newcommand{\dmuA}{\Delta\mutens_{\rm A}}
\newcommand{\dmuB}{\Delta\mutens_{\rm B}}

\newcommand{\depsU}{\Delta\epstens_{\rm U}}
\newcommand{\dmuU}{\Delta\mutens_{\rm U}}

\newcommand\nota[3]{#1_{#2}^{\rm #3}}

\newcommand{\IA}{\mathbbm{1}_{\Omega_{\rm A}}}
\newcommand{\IB}{\mathbbm{1}_{\Omega_{\rm B}}}

\begin{document}

\title{A coupling model for quasi-normal modes of photonic resonators}
\author{Benjamin Vial}
\author{Yang Hao}
\affil{School of Electronic Engineering and Computer Science, 
Queen Mary University of London, 
London E1 4NS, United Kingdom}
\date{\today}

\maketitle

\begin{abstract}
\bfseries We develop a model for the coupling of quasi-normal modes in open photonic systems consisting 
of two resonators. By expressing the modes of the 
coupled system as a linear combination of the modes of the individual particles, we obtain a 
generalized eigenvalue problem involving small size dense matrices. We apply this technique 
to dielectric rod dimmer of rectangular cross section for Transverse Electric (TE) polarization in a two-dimensional (2D) setup.
The results of our model show excellent agreement with full-wave finite element simulations. We provide a convergence analysis, and 
a simplified model with a few modes to study the influence of the relative position of the two resonators. 
This model provides interesting physical insights on the coupling scheme at stake in such systems and pave the 
way for systematic and efficient design and optimization of resonances in more complicated systems,
for applications including sensing, antennae and spectral filtering.
\end{abstract}

\section{Introduction}

Metamaterials are a class of material engineered to produce properties that do not occur naturally \cite{Zheludev2012}. 
Their fundamental building blocks, usually called meta-molecules or meta-atoms, allow to mould the flow of light in 
an unprecedented way \cite{Pendry1780}. The study of these individual building blocks is of prime importance since their shape and 
electromagnetic properties governs the effective behaviour of the metamaterial. In particular, it is of great 
interest to distinguish which effects arise from the meta-atoms themselves or from their arrangement (periodic or random). 
In addition, there is an ever increasing demand for controlling optical fields 
at the nanoscale for applications ranging from medical diagnostics and sensing to optical devices and
optoelectronic circuitry \cite{zeng2014nanomaterials,Singh,Ozbay189,li2008harnessing}. In particular, resonant local field enhancement is of paramount 
importance in phenomena such as surface enhanced Raman scattering (SERS) \cite{SERS2013,stiles2008surface}, improved non linear effects 
\cite{EnhancNonlinear1,novotny2012,nphotonZayats}, optical antennae and the 
control of the local density of states \cite{hoppener2012self,belacel2013controlling} or spectral filtering \cite{vial2014resonant,vial2014transmission}. 
Resonant phenomena are largely at the core of all those applications, calling for a systematic study of modal 
properties of complex nanostructures.\\
Recently, a substantial amount of work have been devoted to the study of resonant interaction 
of light with nanophotonic systems, with particular focus on the computation and analysis of their quasi-normal modes (QNMs \cite{Settimi2009,Lalanne2013}), 
also known as leaky modes \cite{Sammut76}, resonant states \cite{fox1961resonant,Muljarov2010}, quasimodes \cite{lamb1964theory,vial2014quasimodal} or
quasi-guided modes \cite{Tikhodeev2002} in the literature). These modes are an intrinsic 
properties of open resonators, and the associated eigenfrequencies are complex-valued (even for lossless materials), the 
real part giving the resonance frequency and the imaginary part being related to the linewidth of the corresponding resonance. 
A method based on quasi-normal modes expansion has been developed to compute the electromagnetic response of 
open periodic and non-periodic systems to an arbitrary excitation, allowing a simple and intuitive description
of their resonant features \cite{vial2014quasimodal}. 
A similar approach has been employed to define mode 
volumes and revisit the Purcell factor in dispersive nanophotonic resonators \cite{Lalanne2013}, and to compute 
QNMs of perturbed nanoparticles for sensing applications \cite{LalaneSensing}.
Another technique called the Resonant State Expansion (RSE \cite{Muljarov2010,Doost2013}) 
consists in treating the system as a perturbation 
of a canonical problem which spectral elements are known in closed form. 
The idea is to compute these perturbed modes and to use them in the modal decomposition. 
However, those approaches are inherently \emph{perturbative} and often rely on the fact that the considered perturbation 
must be much smaller or with low refractive index compared the unperturbed system.\\
To study the interaction of electromagnetic systems, Coupled mode theories (CMT) \cite{CMT1} have been developed 
using various formulations and extensively applied in
particular in the context of optical waveguides (see Ref. \cite{CMT2} and references therein). 
More recently, it has been extended to leaky eigenmodes with application in 
free space resonant scattering \cite{PhysRevA.75.053801}, 
absorption in semiconductors \cite{Yu12} and Fano resonances in photonic crystal slabs \cite{Fan03}. 
After expanding the modes of the two uncoupled structures onto the modes of isolated simpler systems, one obtains a
system of ordinary differential equations, either in time or space. However, the coupled modes equations are of 
a rather heuristic nature and rely upon a slowly varying approximation, 
thus limiting the domain of application of the method. More recently, hybridization models have been introduced 
in the context of plasmonics \cite{plasmonhybrid}, providing a 
simple and intuitive picture (an electromagnetic analogue of molecular orbital theory) to describe
the plasmon response of complex nanostructures of arbitrary shape as the interaction of plasmons 
of simpler sub-structures. However, the formalism is limited to the quasi-static approximation.\\
The method we develop here is general and relies only on the assumption that the coupled eigenmodes can be expressed as a superposition 
of the modes of the two uncoupled systems, which can be computed by a suitable numerical method or analytically in some simple cases. 
It can be applied to arbitrary shaped open resonators with both non trivial (possibly anisotropic) permittivity and permeability. 
We stress here that it handles \emph{quasi-normal modes} associated with complex eigenvalues. 
The imaginary part of the eigenfrequency giving the leakage rate of the mode is fully taken into account in our method. 
In contrast with the CMT where the coupled modes amplitudes are obtained through solving a system of ordinary differential equations, 
the proposed model reads a generalized eigenvalue problem for the complex coupling coefficients and eigenfrequencies. 
We illustrate the validity of the proposed model on a 2D problem 
of a high index dimmer of rectangular cross section rods for Transverse Electric (TE) polarization. 
The agreement between the model and full-wave finite element simulations is excellent and the convergence is 
analysed as function of the number of basis modes used. Using our model, we then study  the coupling of the two nanorods lowest order modes 
for symmetric and asymmetric dimers by using the first two modes of the isolated resonator. 
This reduced order model depicts a simple hybridization picture, allowing a better understanding of the coupled modes 
in terms of their fundamental spectral building blocks.\\

\section{Coupled mode model}

Consider a system A (material properties $\muA$ and $\epsA$) with eigenvalues 
$\nota{\Lambda}{n}{a}=(\nota{\omega}{n}{a}/c)^2$ and non zero eigenvectors $\nota{\B E}{n}{a}$ 
and a system B (material properties $\muB$ and $\epsB$)
with eigenvalues $\nota{\Lambda}{n}{b}=(\nota{\omega}{n}{b}/c)^2$ and eigenvectors $\nota{\B E}{n}{b}$. 
We study the situation where the two resonators A and B form a coupled system denoted C (material properties $\muC$ and $\epsC$), 
with eigenvalues $\nota{\Lambda}{n}{c}=(\nota{\omega}{n}{c}/c)^2$ and eigenvectors $\nota{\B E}{n}{c}$. 
The spectral equation for the three systems $\rm u\in\{\rm a\,,\rm b\,,\rm c\}$ is:
\begin{equation}
\Mop_{\muU}(\nota{\B E}{n}{u}):=\rot\left[\muU^{-1}\,\rot \nota{\B E}{n}{u}\right]
=\nota{\Lambda}{n}{u} \,\epsU \, \nota{\B E}{n}{u},
\label{eq:eigenpbC}
\end{equation}
In the three cases, the eigenmodes are normalized according to\cite{hanson2002operator}: 
\begin{equation}
 \bra  \epsU \nota{\B E}{n}{u} \mb \nota{\B E}{m}{u}\ket =  
 \int_{\Omega}\epsU(\B r) \nota{\B E}{n}{u}(\B r) \cdotp \nota{\B E}{m}{u}(\B r)\; \ddroit\B r
 =\delta_{nm},
\label{eq:ortho}
\end{equation} 
where $\Omega$ denotes the computational domain.\\
The only assumption we make here is to consider that the coupled eigenmodes can be 
written as a linear combination of the modes of the two isolated systems:
\begin{equation}
 \nota{\B E}{}{c}=\sum_i A_{i} \nota{\B E}{i}{a}  +  B_{i} \nota{\B E}{i}{b}.
 \label{eq:lincomb}
\end{equation} 
Inserting Eq.~(\ref{eq:lincomb}) in the eigenvalue equation (\ref{eq:eigenpbC}), taking into account the 
linearity of operator $\Mop_{\muC}$ and projecting onto eigenmodes $\nota{\B E}{j}{a}$ and $\nota{\B E}{j}{b}$, we 
obtain a generalized eigenvalue problem:
 \begin{equation}
  \B N\,\B C = \nota{\Lambda}{}{c} \B M\,\B C  .
  \label{eq:coupledpb}
 \end{equation}
Here we defined the eigenvectors $\B C=[A_i\,, B_i]^{\rm T}$, $i \in \mathbb{N}$ containing 
the expansion coefficients and the block matrices $\B N$ and $\B M$:
$$\B N=\left(
\begin{array}[c]{cc}
\nota{\B N}{}{aa} & \nota{\B N}{}{ba}\\
\nota{\B N}{}{ab} & \nota{\B N}{}{bb}
\end{array}\right)
\text{,}\quad \B M=\left(
\begin{array}[c]{cc}
\nota{\B M}{}{aa} & \nota{\B M}{}{ba}\\
\nota{\B M}{}{ab} & \nota{\B M}{}{bb}
\end{array}\right),$$
where the coupling coefficients $\nota{N}{ij}{uv}$ and $\nota{M}{ij}{uv}$, $\rm u\in\{\rm a\,,\rm b\}$, 
$\rm v\in\{\rm a\,,\rm b\}$, 
are defined as:
\begin{align*}
\nota{N}{ij}{uv}&=\bra\Mop_{\muC}(\nota{\B E}{i}{u}) \mb \nota{\B E}{j}{v}  \ket , 
&\nota{M}{ij}{uv}&=\bra \epsC \nota{\B E}{i}{u}   \mb  \nota{\B E}{j}{v}  \ket.
\end{align*}
 Thus finding the coupled eigenmodes 
 $\nota{\B E}{}{c}$ (\ie the expansion coefficients $A_i$ and $B_i$) 
 and the coupled eigenvalues $\nota{\Lambda}{}{c}$ boils down to solving the generalized eigenvalue 
 problem (\ref{eq:coupledpb}). In practice we retain $M_{\rm a}$ and $M_{\rm b}$ modes of systems A and B respectively
 in the expansion (\ref{eq:lincomb}), so that $\B N$ and $\B M$ are dense matrices 
 of size $M_{\rm a}+M_{\rm b}$. On the other hand, 
 the matrices involved in FEM based methods are sparse and of generally larger size.\\
 We continue by rewriting the permittivity and inverse permeability of the coupled system as $\epsC=\epsA+\depsB=\epsB+\depsA$ and 
 $\muC^{-1}=\muA^{-1}+\dmuB^{-1}=\muB^{-1}+\dmuA^{-1}$. 
Here we introduced the permittivity and inverse permeability contrasts 
$\depsU$ and $\dmuU^{-1}$ which are bounded by the resonator 
${\Omega_{\rm U}}$, $\rm U\in\{\rm A\,,\rm B\}$:
$$
\begin{array}[c]{cc}
& \depsA=(\epstens_{\rm A}   - \epstens_{\rm h}   )\IA ,\; 
  \depsB=(\epstens_{\rm B}   - \epstens_{\rm h}   )\IB , \\
 & \dmuA^{-1}=(\mutens_{\rm A}^{-1}   - \mu_{\rm h}^{-1}  )\IA,\;
  \dmuB^{-1}=(\mutens_{\rm B}^{-1}   - \mu_{\rm h}^{-1}  ) \IB,
\end{array}
$$ 
where $\IA$ (resp. $\IB$) is the indicator function of domain ${\Omega_{\rm A}}$ (resp. ${\Omega_{\rm B}}$). 
Using the linearity of the involved differential operators and the eigenvalue equations we obtain 
$$ \nota{N}{ij}{uv} = \nota{\Lambda}{i}{u} \nota{L}{ij}{uv}+\nota{K}{ij}{uv} \; 
\text{and} \; \nota{M}{ij}{uv} = \nota{L}{ij}{uv}+\nota{P}{ij}{uv},$$ 
where 
 \begin{equation*}
 \nota{L}{ij}{uv}=\bra \epstens_{\rm u} \nota{\B E}{i}{u}   \mb  \nota{\B E}{j}{v}  \ket=
  \int_{\Omega}\epstens^{\rm u}(\B r) \nota{\B E}{i}{u}(\B r) \cdotp \nota{\B E}{j}{v}(\B r)\; \ddroit\B r,
 \end{equation*}
 \begin{align*}
 \nota{K}{ij}{uv}&=\bra  \Mop_{\Delta\mutens_{\rm V}}(\nota{\B E}{i}{u}) \mb \nota{\B E}{j}{v}  \ket \\
  &=\int_{\Omega_{\rm V}}   \rot\left[ \Delta\mutens_{\rm V}^{-1}(\B r)\,\rot \nota{\B E}{n}{u}(\B r)\right]  \cdotp \nota{\B E}{j}{v}(\B r)\; \ddroit\B r ,
 \end{align*}

  \begin{equation*}
 \nota{P}{ij}{uv}=\bra \Delta\epstens_{\rm V} \nota{\B E}{i}{u}   \mb  \nota{\B E}{j}{v}  \ket
 =  \int_{\Omega_{\rm V}}\Delta\epstens_{\rm V}(\B r) \nota{\B E}{i}{u}(\B r) \cdotp \nota{\B E}{j}{v}(\B r)\; \ddroit\B r.
 \end{equation*}
 Exploiting the orthonormality of the modes, we have 
 $\nota{L}{ij}{aa}=\nota{ L}{ij}{bb}=\delta_{ij}$. We define the block matrices
$$ 
\B\Lambda=\left(
\begin{array}[c]{cc}
\nota{\B\Lambda}{}{a}  & \B{0}^{\rm ba}\\
\B{0}^{\rm ab} & \nota{\B\Lambda}{}{b}
\end{array}\right)
\text{,}\quad  \B L=\left(
\begin{array}[c]{cc}
\B{I}^{\rm a} & \nota{\B L}{}{ba}\\
\nota{\B L}{}{ab} & \B{I}^{\rm b}
\end{array}\right)
$$ 
$$ 
\B K=\left(
\begin{array}[c]{cc}
\nota{\B K}{}{aa} & \nota{\B K}{}{ba}\\
\nota{\B K}{}{ab} & \nota{\B K}{}{bb}
\end{array}\right)
\text{,}\quad \B P=\left(
\begin{array}[c]{cc}
\nota{\B P}{}{aa} & \nota{\B P}{}{ba}\\
\nota{\B P}{}{ab} & \nota{\B P}{}{bb}
\end{array}\right)\text{,}
$$
where $\B{0}^{\rm ba}$ (resp. $\B{0}^{\rm ab}$) is the zero rectangular matrix of size $M_{\rm b}\times M_{\rm a}$ (resp. $M_{\rm a}\times M_{\rm b}$), 
$\B{I}^{\rm a}$ (resp. $\B{I}^{\rm b}$) is the identity matrix of size $M_{\rm a}$ (resp. $M_{\rm b}$), and
$\nota{\B\Lambda}{}{a}$ (resp. $\nota{\B\Lambda}{}{b}$) is a diagonal matrix containing 
the eigenvalues $\nota{\Lambda}{i}{a}$ (resp. $\nota{\Lambda}{i}{b}$). 
We can finally recast the eigenproblem (\ref{eq:coupledpb}) as
 \begin{equation}
  (\B\Lambda\,\B L+\B K)\,\B C = \nota{\Lambda}{}{c} (\B L+\B P)\,\B C.
  \label{eq:coupledpb2}
 \end{equation}
 Writing the eigenvalue problem under this form has two main advantages.
 First, it shows explicitly the relation between the spectrum of the coupled
 system and the spectra of the two uncoupled resonators through the matrix
 $\B \Lambda$. Secondly, the elements of matrices $\B K$ and $\B P$ are computed as overlap
 integrals over a smaller domain, either resonator A or B, thus the numerical 
 calculations are faster and more accurate. \\

 \section{Numerical example}

\subsection{Symmetric dimer}
\begin{figure}[t]
 \centering
 \includegraphics[width=0.9\columnwidth]{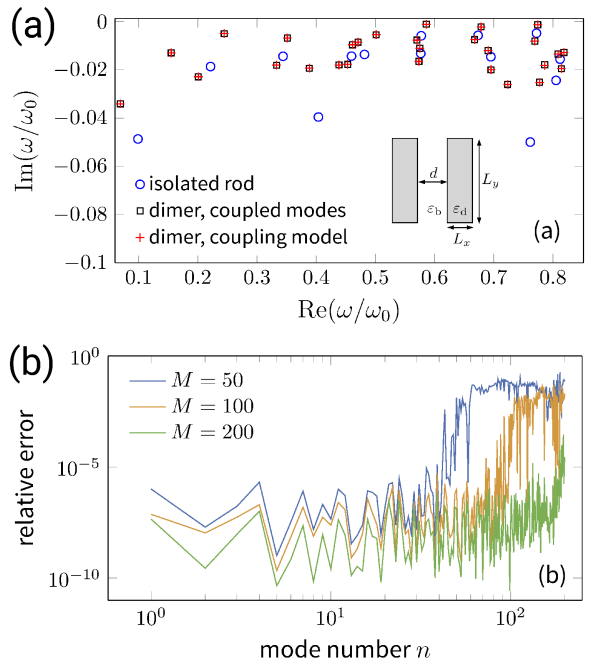}
 \caption{Spectra and convergence of the method. 
 (a): Spectrum for an isolated nanorod (blue circles) and spectra for the dimer 
 computed directly (black squares) and using the coupling proposed model (red crosses, $M=200$). The inset shows 
 a schematic of the dimer and notations. The spectra are normalized with respect to a reference frequency $\omega_0=2\pi c/\L_y$. 
 (b): Relative error of the model $E_{\rm rel}$ (Eq. (\ref{errel})) on the $n$\textsuperscript{th} 
 eigenfrequency for $M=50$ (blue line), $M=100$ (orange line) and 
 $M=200$ (green line) modes retained in the expansion.\label{fig1}}
\end{figure}

To illustrate the method we consider a 2D example for the TE polarization case, \ie~$\B E=E_z\B e_z$. 
The structure is dimer consisting of two identical 
dielectric rods ($\varepsilon=16$, $\mu=1$) of rectangular 
cross section ($L_x=50\,$nm, $L_y=3\,L_x=150\,$nm) embedded in air and separated by a gap $d=1.4L_x=70\,$nm 
(cf. inset in Fig.~\ref{fig1} (a)). 
The modes of the individual rod and of the dimer are computed with a Finite Element Method (FEM) 
using Perfectly Matched Layers (PMLs) to truncate the infinite air domain \cite{vial2014quasimodal}.\\
We first study the convergence of the model as a function of the number
of modes $M=M_{\rm a}=M_{\rm b}$ retained in the expansion ($M= 50$, $100$ or $200$) for the first $200$ modes 
(ordered with increasing real parts of eigenfrequency). 
Results are plotted on Fig.~\ref{fig1} (a) for $M=200$, where we can see that each eigenfrequency $\omega$ of 
an isolated rod produce a pair of coupled frequencies $\nota{\omega}{\pm}{c}$ corresponding to a bright  
and dark mode. Our model can predict the position of these coupled eigenfrequencies in the complex plane 
with very good accuracy. To quantify this we computed the relative error for the eigenfrequencies 
calculated with our model (denoted $\nota{\omega}{n}{c}$) and the reference values for the dimer computed 
directly with the FEM (denoted $\nota{\omega}{n}{c,\,ref}$) defined as 
\begin{equation}
 E_{\rm rel}=\left|\frac{\nota{\omega}{n}{c}-\nota{\omega}{n}{c,\,ref}}{\nota{\omega}{n}{c,\,ref}}\right| .
 \label{errel}
\end{equation}
We can see on Fig.~\ref{fig1} (b) that the relative error is between $10^{-6}$ and $10^{-10}$ for $M=200$. 
Taking more basis modes into account improves the overall convergence, particularly 
for higher order modes, as can be seen from the relative error curves for 
$M=50$, $100$ and $200$. This proves the convergence of the method and its ability 
to compute the spectrum of the coupled system with high accuracy. Note that here we included proper modes (\ie quasi-normal modes associated 
with point spectrum) as well as improper modes 
(radiation modes associated with continuous spectrum, dicretized by using PMLs \cite{olyslager2004discretization}). From a practical point of view, 
the computational time increases as $M^2$ with $t_{\rm c}=77$s, $297$s and $1233$s for $M=50$, $100$ and $200$ respectively.\\

\begin{figure}[t]
 \centering
 \includegraphics[width=0.9\columnwidth]{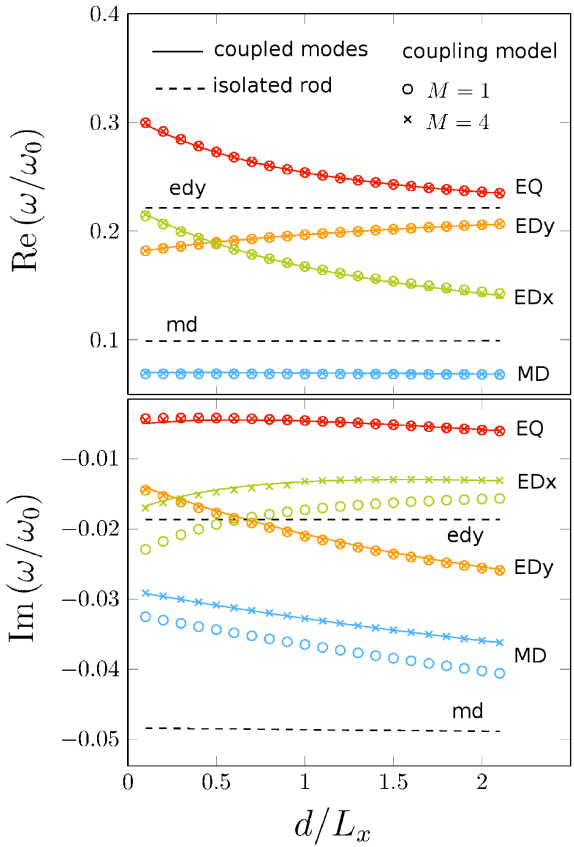}
 \caption{Coupled modes of a symmetric dimer. Real (top) and imaginary (bottom) parts of the eigenfrequencies 
 as a function of the normalized gap $d/L_x$ for the first four coupled modes: 
 the magnetic dipole (MD, blue), the electric dipole along $x$ 
 (EDx, green), the electric dipole along $y$ (EDy, orange) and the electric quadrupole (EQ, red). 
 The coupled eigenfrequencies computed using our model with $M=1$ (circles) and $M=4$ (crosses) agree 
 well with the coupled eigenfrequencies computed directly (solid lines). The first two modes of the isolated rod,
 the magnetic dipole (md) and the electric dipole along $y$ (edy), are indicated with horizontal dashed lines.
\label{fig2}}
\end{figure}

\begin{figure}
 \centering
 \includegraphics[width=0.9\columnwidth]{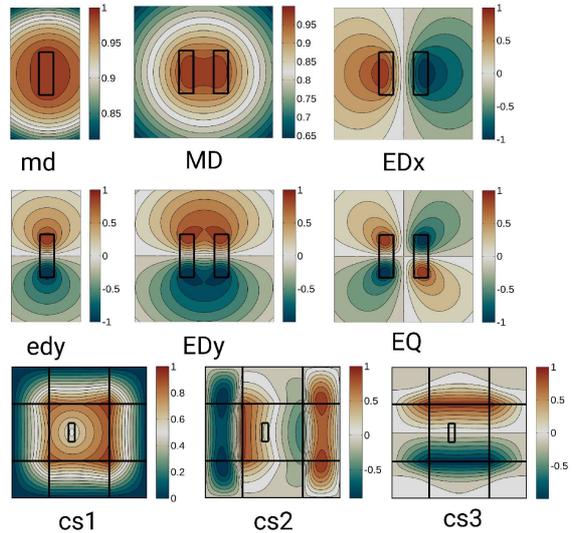}
 \caption{Electric field maps of the modes studied for the isolated rod (md, edy, and modes of the continuous spectrum cs1, cs2 and cs3) 
 and coupled modes of the dimer (MD, EDx, EDy, EQ) for $d/L_x=1.4$.  \label{fig3}}
\end{figure}

To gain insight into the coupling mechanism between
different eigenstates, we restrain the number of basis elements included in our model. 
We study the first four coupled modes (\ie with the lowest real parts of their eigenfrequency) 
as a function of the gap of the dimer $d$. 
The traditional hybridization picture can be seen as a particular case of our model 
where one retains only one QNM in the expansion. Indeed, one mode usually dominates 
the coupling process for the low frequency modes, but this is not true for higher 
order modes. In the particular case studied here, we use only the 
two lowest order modes of the isolated rod : the magnetic dipole (denoted md) and 
the electric dipole along $y$ (denoted edy) (cf. Fig.~\ref{fig3}). 
Symmetry considerations prove that 
these two modes do not couple to each other, so we can use a one mode 
model to find the coupled resonant frequencies of the first four modes of the dimer. In the case of non 
magnetic materials ($\mu=1$) we have $K=0$ and because the system is symmetric with respect 
to the $x=0$ plane, we have $L^{ab}=L^{ba}=L$, $P^{aa}=P^{bb}=P'$, $P^{ab}=P^{ba}=P''$, 
and $\Lambda_a=\Lambda_b=\Lambda$. Under these conditions we obtain a 
simple analytical formula for the coupled eigenfrequencies:
\begin{equation}
\nota{\omega}{\pm}{c}=  \sqrt{\frac{ 1\pm L }{1 \pm L + P' \pm P''} }\,\omega.
\label{onemodeapprox}
\end{equation}
The associated eigenmodes are $C_+=(1,1)$ which represent an in phase coupling (the so-called bright mode), 
and $C_-=(1,-1)$ accounting for an out of phase coupling (dark mode). In our case, the magnetic dipole (md) modes 
of the two isolated particles give rise to the magnetic dipole (MD) and the electric dipole along $x$ (EDx) 
of the dimer thanks to a symmetric and antisymmetric coupling respectively (cf. Fig.~\ref{fig3}). Similarly, 
the in-phase (resp. out-of phase) coupling of the two electric dipole modes along $y$ creates the coupled electric dipole mode along $y$ EDy
(resp. the coupled electric quadrupole mode EQ). This single mode approximation 
is clearly valid for the particular setup used here as can be seen from Fig.~\ref{fig2}, 
where the results from the one mode model (circles) are really close to the 
reference coupled eigenfrequencies computed directly (solid lines). However Eq.~(\ref{onemodeapprox}) is less accurate for
the imaginary part of the coupled eigenvalues in the case of the coupling of two magnetic dipoles 
(cf. blue and green curves for MD and EDx). We thus included three modes of the continuous spectrum in the model (denoted cs1, cs2 and cs3 
with eigenvalues $\omega_{\rm cs1}=(0.0481 - 0.0726\ic)\,\omega_0$, $\omega_{\rm cs2}=(0.0593 - 0.1080\ic)\,\omega_0$ 
and $\omega_{\rm cs3}=(0.0594 - 0.1090\ic)\,\omega_0$), which improved greatly the accuracy on the imaginary part 
(cf. crosses on Fig.~\ref{fig2}). This result suggest that one need to include radiation modes in the model 
to account properly for the leakage of the coupled modes. We have checked that the accuracy does not improve if 
we add higher order QNMs for both real and imaginary parts. On the contrary, the accuracy increases when we add modes 
from the continuous spectrum. It is not clear from a mathematical point of view whether the QNMs 
form a complete basis outside the resonator in the general case as to the best of our knowledge it
has been proved to be valid only inside the resonator 
and only for one-dimensional systems \cite{PhysRevA.49.3982}. However, there is numerical evidence 
that QNMs and radiation modes are \emph{both} needed in modal expansion for computing accurately electromagnetic fields \cite{vial2014quasimodal}. 
From a physical point of view, our interpretation is that including modes from the continuous spectrum helps reconstructing 
the eigenfield \emph{outside} of the resonator, describing more accurately its radiative nature and thus increasing 
the precision on the imaginary part of the eigenfrequency, which reflects the leakage rate of the mode.\\

\subsection{Breaking the symmetry}

\begin{figure}
 \centering
\includegraphics[width=0.9\columnwidth]{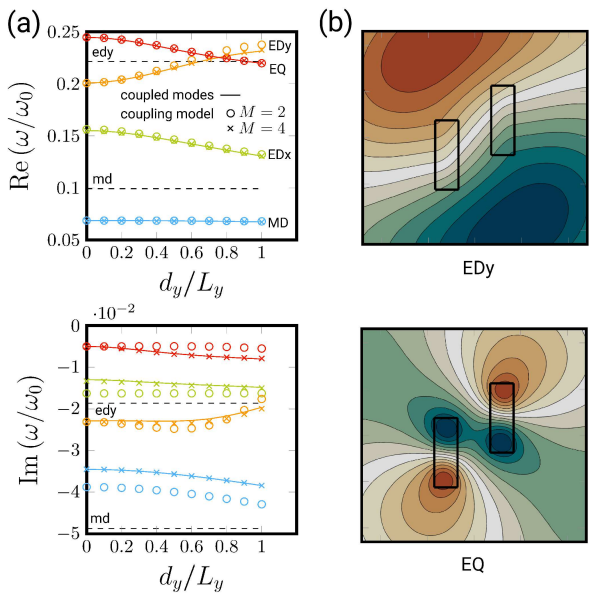}
 \caption{Coupled modes of an asymmetric dimer. (a): Real (top) and imaginary (bottom) parts of the eigenfrequencies for the first four 
 coupled modes as a function of normalized shift $d_y/L_y$ ($d/L_x=1.4$). Solid lines: coupled modes (full wave direct computation), 
 circles: two modes coupling model, crosses: four modes coupling model. (b): Electric dipole along $y$ and electric quadrupole modes for $d_y/L_y=0.5$. 
\label{figshift1}}
 \end{figure}

\begin{figure}
 \centering
\includegraphics[width=0.9\columnwidth]{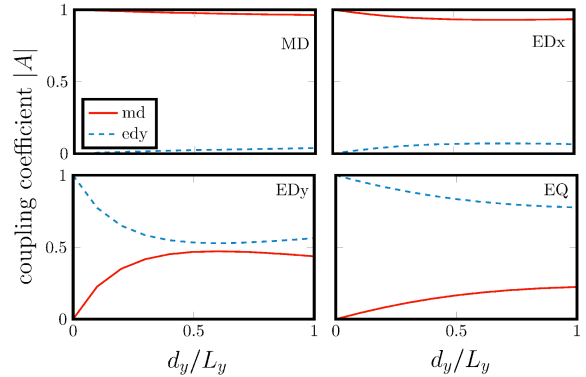}
 \caption{Norm of the coupling coefficient $|A|$ as a function of the 
 normalized shift $d_y/L_y$ for the four coupled modes.\label{figshift2}}
 \end{figure}

Finally we study the case where one of the resonator is shifted of a distance $d_y$ along $y$, 
so that the symmetry of the original dimer is broken. The two fundamental modes of the 
dielectric rods (md and edy) can now couple to each other, so we include them both in our model. 
Results are reported on Fig.~\ref{figshift1} (circles) and show good agreement with full wave simulations (full lines). 
When the shift increases, the coupled magnetic dipole resonance barely changes, EDx and EQ frequencies are redshifted while EDy is blueshifted. 
The prediction of the model on the imaginary part is correct but less precise. Adding the radiation modes cs1 and cs2 
in the expansion reduces those discrepancies significantly (see crosses on Fig.~\ref{figshift1} for $M=4$).\\
The study of the coupling coefficients allow a quantitative evaluation 
of the coupling between modes induced by the breaking of symmetry. As the two resonators are identical, 
we have for a given coupled mode $|A_n|=|B_n|$ ($n$ being the index of the isolated mode, either md or edy),
with a relative phase difference of $0$ for the symmetric coupling 
and $\pi/2$ for the antisymmetric coupling. We plot those coefficients as a function 
of the shift for the four coupled modes on Fig.~\ref{figshift2}. For coupled 
MD and EDx, the md still dominates the coupling, either symmetrically or antisymmetrically, which is 
illustrated by the fact that the norm of the related coupling coefficient 
$|A|$ associated with md (red solid line) is greater than the one for edy. 
The situation is more complicated for the other two modes: for EDy, increasing the shift 
creates more symmetric coupling between the two isolated modes, resulting in a 
hybrid mode where the contribution of md and edy is almost the same for large shifts 
(cf. the bottom left panel on Fig.~\ref{figshift2}, where 
$|A_{\rm md}|\simeq |A_{\rm edx}|$ for $d_y/L_y=0.5$ and the associated field map on Fig.~\ref{figshift1}~(b)). 
The situation is similar for the EQ mode with an antisymmetric coupling of both nanorod modes, with 
an increasing contribution of the magnetic dipole as $d_y$ increases (cf. the bottom right panel on Fig.~\ref{figshift2}
and field map on Fig.~\ref{figshift1}~(b)).

 \section{Conclusion}
We have developed a coupling model for computing the eigenmodes of two coupled electromagnetic resonators. 
Its validity is illustrated in the 2D TE case of high dielectric object through its good agreement
in comparison with full wave simulations. The vectorial formulation is general and is suitable for arbitrarily shaped 
finite resonators with complex permittivity and permeability (possibly spatially varying and anisotropic). 
Future research directions will focus on applying the model to the Transverse Magnetic (TM) case and three dimensional problems, 
as well as extending it to periodic systems. 
Note that our approach can be straightforwardly extended to the coupling 
of an arbitrary number of objects by simply adding more terms corresponding to the modes 
of the different objects in Eq.~(\ref{eq:lincomb}), allowing the study of photonic oligomers \cite{Oligomer}. 
This model provides physical insights on the coupling of individual modes and may ease the 
design and understanding of resonant processes required for many applications in photonics.

\section{Acknowledgments}
This work has been funded by the Engineering and Physical
Sciences Research Council (EPSRC), UK under a
Programme Grant (EP/I034548/1) ‘‘The Quest for
Ultimate Electromagnetics using Spatial Transformations
(QUEST).’’


\end{document}